\documentclass[preprint2, times]{aastex631}
\let\tablenum\relax

\usepackage{booktabs, siunitx, stix}
\renewcommand\micron{\textmu{m}}

\shorttitle{JWST Spectrophotometry of the Small Satellites of Uranus and Neptune}

\begin{document}

\title{JWST Spectrophotometry of the Small Satellites of Uranus and Neptune}
\correspondingauthor{Matthew Belyakov}
\email{mattbel@caltech.edu}
\author[0000-0003-4778-6170]{Matthew Belyakov}
\affiliation{Division of Geological and Planetary Sciences, California Institute of Technology, Pasadena, CA 91125, USA}

\author[0000-0002-7451-4704]{M. Ryleigh Davis}
\affiliation{Division of Geological and Planetary Sciences, California Institute of Technology, Pasadena, CA 91125, USA}

\author[0000-0001-5683-0095]{Zachariah Milby}
\affiliation{Division of Geological and Planetary Sciences, California Institute of Technology, Pasadena, CA 91125, USA}

\author[0000-0001-9665-8429]{Ian Wong}
\affiliation{NASA Goddard Space Flight Center, 8800 Greenbelt Road, Greenbelt, MD 20771, USA}

\author[0000-0002-8255-0545]{Michael E. Brown}
\affiliation{Division of Geological and Planetary Sciences, California Institute of Technology, Pasadena, CA 91125, USA}

\begin{abstract}
We use 1.4-4.6 \micron{} multi-band photometry of the small inner Uranian and Neptunian satellites obtained with the James Webb Space Telescope's near-infrared imager NIRCam to characterize their surface compositions. We find that the satellites of the ice giants have, to first-order, similar compositions to one another, with a 3.0 \micron{} absorption feature possibly associated with an O-H stretch, indicative of water ice or hydrated minerals. Additionally, the spectrophotometry for the small ice giant satellites matches spectra of some Neptune Trojans and excited Kuiper belt objects, suggesting shared properties. Future spectroscopy of these small satellites is necessary to identify and better constrain their specific surface compositions.
\end{abstract}

\section{Introduction} \label{sec:intro}
Thirty-five years after Voyager 2's fly-by of Uranus and Neptune \citep{1986Sci...233...43S, 1989Sci...246.1422S}, the satellites of the ice giants remain understudied as compared to those of the gas giant systems. The lack of dedicated spacecraft exploration, combined with the difficulty of ground-based observations due to intense scattered light from the planets, has left the innermost satellites of the ice giants without spectral characterization of their surfaces. Much of what is known about these satellites comes from Voyager 2 images, along with HST and ground-based visible and near-IR photometry \citep{1989Sci...246.1422S, 1992A&A...262L..13C, 2001Icar..151...51K, 2002AJ....123.1776D, 2003Icar..162..400K, pascu2006hst}. 

Constant cycling of material from collisions and subsequent re-accretion on million year timescales is the dominant process shaping the small interior satellites of the ice giants \citep{1992JGR....9710227C, french2012cupid, 2018prs..book..517C, 2022AJ....164...38C}. Catastrophic collisions to these moons from Kuiper belt objects and Oort Cloud comets are less frequent, occurring on 2 Gyr timescales for the smallest moons, and on timescales significantly longer than the lifetime of the solar system for the larger interior satellites like Neptune's Proteus or Uranus' Puck \citep{zahnle2003cratering, 2022AJ....164...38C}. While the larger satellites of Uranus (Miranda, Ariel, Umbriel, Titania, Oberon) and Neptune (Triton) all present their own interesting geology \citep{croft1991geology, 1995netr.conf..879C, cartwright2015distribution,2020RSPTA.37800102S, 2020ApJ...898L..22C, 2021PSJ.....2..137H}, the heavy endogenous modification of their surfaces makes them poor probes of the original composition of the ice giants and their subnebulae. Although the exact histories of the inner satellites of Uranus and Neptune are unknown, they are likely a combination of ice giant subnebulae material and interloping comets, with larger objects being close to the original subnebula material, and smaller satellites being of mixed composition \citep{1992JGR....9710227C, szulagyi2018situ,2019AJ....157...30H, 2022AJ....164...38C}.

\begin{figure*}
    \centering
    \includegraphics{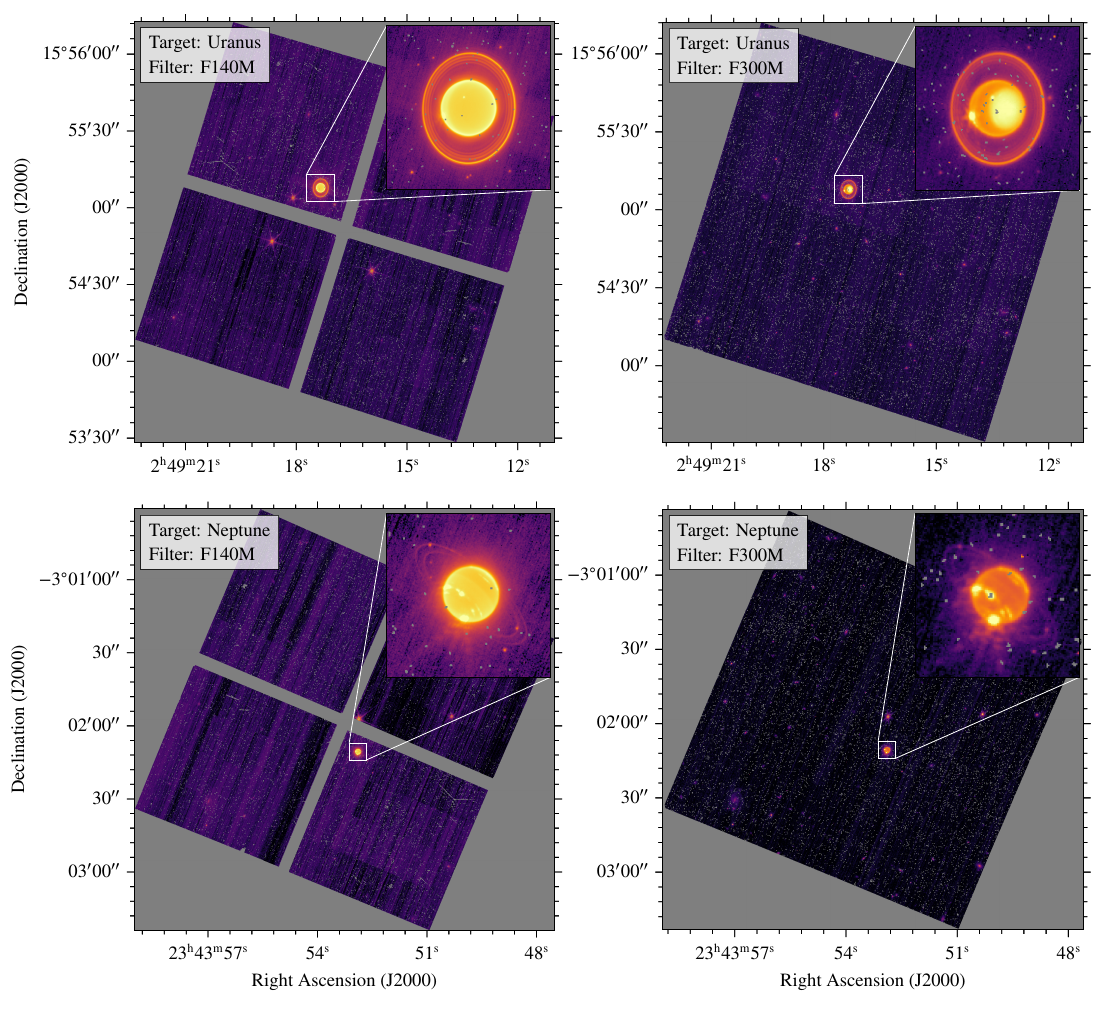}
    \caption{Representative JWST NIRCam images (brightness is square root scale) of the Neptunian and Uranian systems in the F140M and F300M filters.}
    \label{fig:reprojected-combined}
\end{figure*}

\begin{figure*}
    \centering
    \includegraphics[width=\textwidth]{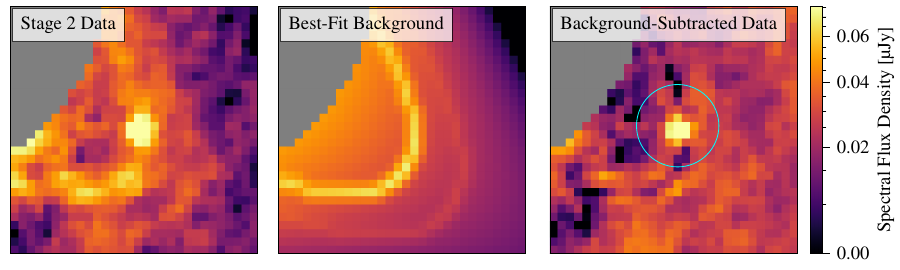}
    \includegraphics[width=\textwidth]{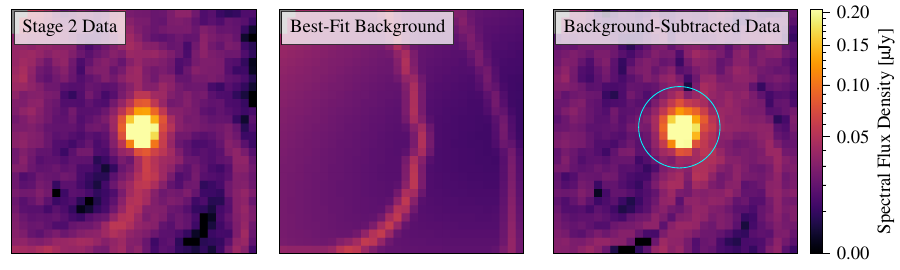}
    \caption{Background subtraction for an F140M (top) and an F300M (bottom) image of Neptune's moon Despina. Despina is situated adjacent to the Le Verrier ring, which has a significant contribution to the flux around the moon. The left panels show the data after stage 2 pipeline reduction which produces absolute flux-calibrated images. The middle panels show the composite background made by combining a model for the rings with a second-degree polynomial surface that characterizes the scattered light background from Neptune The background-subtracted data is on the right, showing effective subtraction of the scattered light gradient and the ring contribution. Note that while each image is $30\times 30$ pixels, the pixel scale in the F300M image is twice that of the F140M image, which accounts for the difference in the apparent shape of the rings and the masking of Neptune in the F300M data.}
    \label{fig:fig1}
\end{figure*}
We briefly review the main properties of the thirteen inner moons of the Uranian system. Although Uranus has a larger and more varied set of rings than Neptune, it has fewer shepherd moons (at least known moons, see \cite{2016AJ....152..211C} for a discussion of the possibility of new Uranian moons) for its rings, with the innermost Cordelia and Ophelia shepherding the $\varepsilon$ and $\lambda$ rings \citep{karkoschka2003sizes, showalter2006second}. The next nine moons in the set (Bianca, Cressida, Desdemona, Juliet, Portia, Rosalind, Cupid, Belinda, Perdita) all share common albedos and spectral slopes with the second-largest moon, Portia, and are collectively described as the ``Portia Group'' \citep{karkoschka2003sizes}. The largest inner moon Puck has a slightly larger visual albedo, and has a distinct spectral slope in the visible from the closer-in satellites \citep{karkoschka2003sizes, 2003AJ....126.1080D}. Given the large number of moons, the system is unstable on more than hundred million year timescales, with collisions and re-accretions expected \citep{french2012cupid,2022AJ....164...38C}.

The inner component of the Neptunian system consists of seven moons, which we briefly review, starting from innermost Naiad and Thalassa. Naiad is locked into a 73:69 resonance with Thalassa \citep{2020Icar..33813462B}, and both moons are likely rubble piles based on their apparently low densities \citep{1989Sci...246.1422S, 1991Icar...93..379G, 2003Icar..162..400K}. The next two moons are Despina and Galatea, which are shepherd moons of the Le Verrier and Adams rings respectively \citep{1991Sci...253..995P, 2002Natur.417...45N}. The rapid evolution of the ring arcs in the Adams ring on decade-long timescales suggest that the orbits of the inner Neptunian moons are still evolving \citep{2014A&A...563A.133R}. Larissa is slightly larger, at 200 km, and orbits just outside Neptune's rings \citep{1989Sci...246.1422S}. The next moon, Hippocamp, was only discovered in 2013, and is likely a small collisional product of the largest of the inner moons, Proteus \citep{2019Natur.566..350S}. Proteus has a notable leading/trailing brightness asymmetry in the visible, attributed to charged-particles from Neptune's magnetosphere bombarding Proteus' trailing hemisphere \citep{pascu2006hst}. However, the source of the asymmetry is unclear, and details about Proteus' surface composition remain unclear. All of the inner moons have similarly low visual albedoes at under 10\% \citep{2003Icar..162..400K} and similar neutral colors \citep{1989Sci...246.1422S}. 

In this paper we use JWST NIRCam four band (1.4, 2.1, 3.0, 4.6 \micron{}) near-IR photometry of the inner moons of Uranus and Neptune obtained via images of the ice giants taken in Program 2739 (JWST Cycle 1 Outreach Campaign). We first discuss the observations and data reduction, focusing on the background subtraction which, due to the complexity of the scattered light from the planets and their rings, has to be handled separately from the standard JWST pipieline. With the resulting spectrophotometry, we compare the satellites with one another and to spectra of other solar system small bodies.

\section{Observations}\label{sec:methods}

The Neptunian and Uranian systems were observed through JWST Program 2739 (JWST Cycle 1 Outreach Campaign) on three separate dates using the near-infrared imager NIRCam. The Neptunian observations were taken on 2022 July 12 when the planet was at a phase angle of 1.8$^\circ$, and are in four filters: F140M, F210M, F300M, and F460M, which are correspondingly centered at 1.4, 2.1, 3.0, and 4.6 \micron{}. The Uranian observations are split between two dates, one set of images taken only in F140M and F300M filters on 2023 February 6 (phase angle 2.9$^\circ$), and a second group of images taken on 2023 September 4 (phase angle 2.8$^\circ$) in the same four filters as the Neptune images. In total, there are 40 images for Neptune and 48 images for Uranus, split between the four filters used. 

Images were taken using NIRCam's module B, which offers higher long-wavelength throughput than module A (NIRCam has two modules, offering slightly different performance; \citealt{2023PASP..135d8001R}). Ten exposures were taken in each filter resulting from five dithers performed with two different exposure strategies. For the Neptune and second set of Uranus images, a first series of exposures was done with an 80~s exposure time (1 integrations, 8 groups, \texttt{RAPID} readout mode), and a second was taken with a 280~s effective exposure time (2 integrations, 7 groups, \texttt{Bright1} readout mode). The short wavelength images are split between four detectors with a 5$^{\prime\prime}$ gap, and the positioning of the planets on the detector frequently results in moons falling on the detector gap or too close to the detector edge in one or more dithers. The choice of dither put Neptune's moon Galatea in the detector gap for three out of five of the F140M and F210M images, while for Uranus, Juliet and Portia are off the detector in the final F210M image.

\section{Data Reduction}\label{sec:photom}
For each image, we begin with the Level 2 ``rate'' data products, which have been bias- and dark-corrected and converted into units of counts/second by the ramp fitting step (JWST pipeline version 1.11.3, \cite{bushouse_2023_8157276}). We then skip the background subtraction of stage two processing, as we apply our own background subtraction scheme later, and proceed with the \texttt{assign\_wcs}, \texttt{flat\_field}, and \texttt{photom} steps, the latter of which converts the ramp files to flux units of MJy/sr. The product of these steps can be seen in \autoref{fig:reprojected-combined}.

At this step, we also discard all of the images taken with the \texttt{Bright1} readout pattern. Program 2739 was designed to provide images of the Neptunian and Uranian systems at large for public release, and thus JWST's tracking rate was set to match the planet's motion, not that of any given satellite. As the moons have an extra on-sky motion of 0.01-0.02 ''/minute (just under the pixel scale of the short wavelength detector) relative to their host planets, they are smeared on the detector. However, the ramp-fitting step in the JWST pipeline encounters issues, especially with the Uranian satellites, as the brightness in the pixels changes as the moon moves over the course of one integration. The sudden drop or jump in counts prevents the ramp-fitting or jump detection steps from accurately fitting the count rate of the object in the longer exposure images. We therefore discard the longer 280~s observations for both the Neptunian and Uranian system, instead using only the 80s exposures, during which the motion of the satellites' PSFs on the detector is on a sub-pixel scale.  

\subsection{Astrometry}
To simplify the analysis process and the mapping between sky coordinates and pixel coordinates, we re-projected each image into rectangular sky coordinates with right ascension increasing to the left along the horizontal axis and declination increasing upward along the vertical axis. For this reprojection, we used the Astropy-affiliated package \texttt{reproject}\footnote{\url{https://reproject.readthedocs.io/en/stable/}}. We first find the optimal WCS for our set of images using the Astropy-affiliated package \texttt{reproject}, then use its flux-conserving interpolation algorithm \texttt{reproject\_adaptive} to create a composite image of all detectors (four detector images for the F140M and F300M data, one detector image for the F210M and F460M data) reprojected into rectangular right ascension and declination sky coordinates (see figure \ref{fig:reprojected-combined} for an example of these reprojections). For each satellite target, we calculated the position in RA/Dec at the time of the observation using JPL Horizons \footnote{\url{https://ssd.jpl.nasa.gov/horizons/}}, which we then mapped to pixel positions using WCS transforms. However, the satellite positions were often different by up to 3 pixels along either axis. The calculated position of Neptune was always accurate, so we suspect the error is due to imprecision in the satellite ephemeris used by Horizons. To calculate the true position of the satellite, we extracted a 10-pixel-by-10-pixel image around the Horizons coordinates, then fit a two-dimensional Gaussian profile with a background constant, which provided us with refined pixel coordinates.

\subsection{Background Subtraction}

Given the proximity of the inner moons of Uranus and Neptune to the planet and the presence of multiple rings, the standard method of taking the median brightness value in an annulus around each satellite is not sufficient to account for the background. Instead, we create a background subtraction that accounts for the scattered light from the planet and any contribution from the rings. 

\begin{figure*}
    \centering
    \includegraphics[width = 8.9cm]{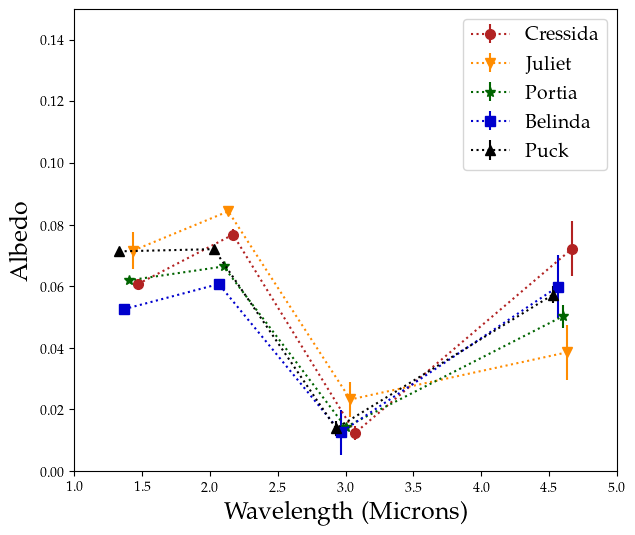}
    \includegraphics[width = 8.9cm]{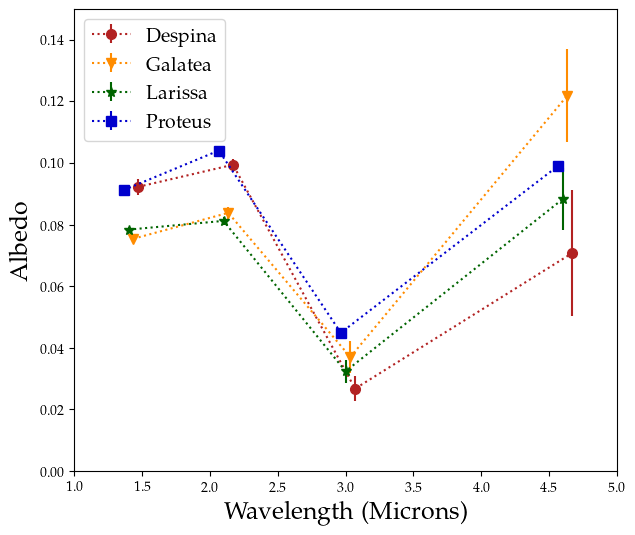}
    \caption{\textit{Left panel:} Albedos of the  $r > 40$ km satellites of Uranus (Cressida, Juliet, Portia, Belinda, and Puck). \textit{Right panel:} Albedos of four inner Neptunian satellites (Despina, Galatea, Larissa, and Proteus) for the four wavelengths observed. Each moon's observations have a slight horizontal offset from the filter's central wavelength to assist readability. Both diagrams share the same vertical axis for easier comparison. All of the satellites show a clear 3 \micron{} absorption feature, and most have an upwards slope between 1.4 and 2.1 \micron{}. The Uranian satellites are on average lower albedo than the Neptunian ones, and seem to have a somewhat smaller F300M-F460M slope.}
    \label{fig:photom}
\end{figure*}

Removing the contribution from the rings is especially important for the satellites of Neptune, where Galatea blends with the Adams ring and Despina blends with the Le Verrier ring. In order to remove the ring flux, we first mask out the targeted satellite, the two rings, and Neptune, then fit a second-degree two-dimensional polynomial surface to the scattered light background. After removing the scattered light contribution, we construct and fit a model of the rings projected into the viewing geometry of the individual observation and the pixel scale of the detector. We begin by constructing a weighted brightness map of the rings using the following procedure. First, for both of the rings, we generate 3,600 separate on-sky coordinates in RA/Dec, azimuthally equidistant in $0.1^\circ$ increments. These RA/Dec positions on the ring are obtained by adding a radial offset in the ring plane of Neptune using the ring radii given in \cite{2018prs..book..112D} to the position of Neptune in JPL Horizons. We then bin the coordinates into a 2D histogram with boundaries defined by the NIRCam pixel edges. This effectively adds up, for each detector pixel, the contribution of each 0.1$\circ$ segment of the ring to the brightness in that pixel. This model therefore determines which pixels encompass a larger section of the ring; e.g., pixels next to the ring ansa are proportionally brighter than other pixels, as they contain more of the ring flux. As a result, we have a map for the relative flux of each pixel in the ring as compared to any other pixel in the ring. We then convolve the histogram with a 2D Gaussian kernel approximating the appropriate filter PSF. Having obtained the relative proportions for the brightness of each component of the ring, we now fit the model ring image to the data for the ring using a simple linear fit using least-squares minimization: we find which coefficient times the ring model plus some constant offset best matches the data. In order to prevent the fit from being skewed by the brightest pixels next to Neptune, we only fit using those pixels in the ring which are 7-12 pixels away from our target of interest. We then subtract the best-fit final ring model from the science image. 

Figure \ref{fig:fig1} shows an example of this method applied to Despina for both the higher spatial-resolution short wavelength mosaic data (top row) and lower spatial-resolution long wavelength single-detector data (bottom row). The first column shows the product of stage 2 pipeline-processed data in units of \textmu Jy (after the \texttt{photom} step of the JWST pipeline). The middle column shows the best fit background composed of the scattered light polynomial surface and the two rings. The right column shows the background-subtracted data, demonstrating the effective removal of the rings without significant residuals.

The background subtraction for Uranus is simpler as all satellites we analyze are exterior to the orbit of the bright $\epsilon$ ring, rather than co-incident with it at the detector scale of NIRCam. However, because some satellites are still affected by scattered light from the bright ring, the rings must be removed before fitting for the scattered light background, and so we simply mask the $\epsilon$ ring completely for all stages of the background subtraction.

\subsection{Photometry}
\begin{figure*}
    \centering
    \includegraphics[width=\textwidth]{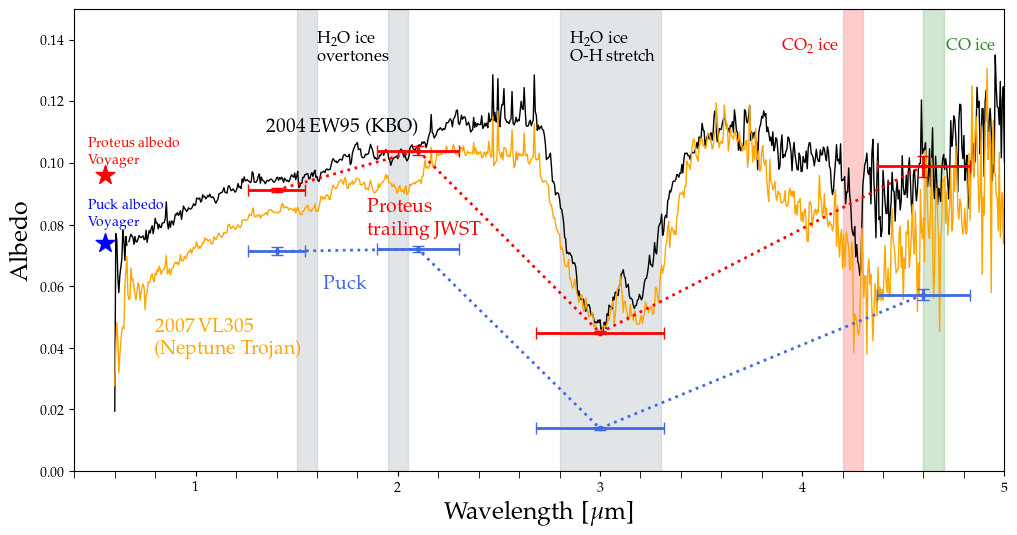}
    \caption{JWST NIRCam albedos of Proteus and Puck compared to the spectra of 2004 EW95 (Plutino) and 2007 VL305 (Neptune Trojan) from JWST Programs 2418 and 2550 respectively. Also shown are Voyager-derived visual geometric albedos, with the Puck albedo selected to be at the same phase angle as the JWST observations (see Figure 6 of \cite{2001Icar..151...51K} and Table 2 of \cite{karkoschka2003sizes}). The spectrophotometry for Proteus and Puck closely matches the shape and albedo for the selected objects, both of which have less red/neutral spectral slopes, and perihelia also near 30AU. Higher resolution spectra for Puck and Proteus are needed to reveal surface composition.}
    \label{fig:compare}
\end{figure*}
In order to obtain accurate photometry, we empirically determine what fraction of light is captured by our chosen aperture radius of six pixels. First, we take the CALSPEC spectrum of solar analog P330E (which is a composite of HST spectra and stellar models) and convolve it with the four filter bandpasses we use in order to obtain synthetic photometry for the star \citep{2014PASP..126..711B, 2015AJ....149..122B, Gordon_2022}. Then, using NIRCam images of the same star from Program 1538 (Absolute Flux Calibration (G Dwarfs)), we draw successively larger apertures, determining what portion of the flux we obtain from photometry relative to the synthetic flux from the spectrum\footnote{We note that comparing to the NIRSpec spectrum is a necessary step, as the internal flux calibration used to generate the counts/s to MJy/sr conversion in the \texttt{photom} step is based off specific choices of aperture size, and it is such that the true flux is not obtained with an infinite aperture in the NIRCam images.}. At six pixels, this is approximately 83\% of the light for F140M, F210M, and F460M, and around 86\% of the light for F300M (with less than 1\% error). We also compare these values to the theoretical PSFs, finding that the curves for the percent of captured light as a function of aperture size closely match those from WebbPSF \footnote{See figures 8/9 of the JDox for JWST NIRCam for an example of encircled energy diagrams \url{https://jwst-docs.stsci.edu/jwst-near-infrared-camera/nircam-performance/nircam-point-spread-functions}} \citep{2016jdox.rept......}. All of our photometric measurements are divided by the proportion of captured flux in the corresponding filter at six pixels to obtain the true value for the flux. 

We convert the intensities obtained from the photometry into geometric albedos using the measured solar flux in the NIRCam filters as given in \cite{2018ApJS..236...47W} along with sizes for the moons obtained by Voyager 2 \citep{1986Sci...233...43S, 1989Sci...246.1422S, 2001Icar..151...51K, karkoschka2003sizes}. We report our photometry in both flux units and as albedos in every band in \autoref{tab:table1}. To obtain an estimate for the error on each flux measurement, we calculate the flux in eight equidistant apertures centered ten pixels from the target, and take the standard deviation of their flux as an estimation of the systematic error present in the image after background subtraction. As there is significant local variation in the background of the images, both due to detector issues and structure from the rings and the planet, calculating the standard deviation of flux in nearby apertures provides a good estimate of the true uncertainty on the measurement. The albedo and flux for each target is then reported as the mean measurement, with the errors from all the images in a given filter added in quadrature, shown in \autoref{fig:photom}.

\section{Discussion}\label{sec:results}
Results from JWST have given extensive insight into the compositions of small bodies in the outer solar system, with CO$_2$, CO, H$_2$O, and CH$_3$OH ices being commonplace on Kuiper belt objects \citep{2023PSJ.....4..130B, 2023arXiv231003998M, licandropreprint}. By comparison, little is known about the composition of the inner satellites of the ice giants. In the case of Neptune, the violent capture of Triton \citep{agnor2006neptune} leaves open the question of what composes the collisional debris that went on to form the satellites interior to Triton \citep{goldreich1989neptune,2011Icar..214..113N, 2020Icar..33813462B}. For the Uranian satellites, the reason for the stark difference between the very dark surfaces of Puck and Portia group satellites as compared to the brighter regular satellites \citep{2001Icar..151...51K} remains unclear.

Our results for the photometry of the inner Uranian and Neptunian satellites (as shown in \autoref{fig:photom}) demonstrate that to first order, the small interior moons seem to all have similar spectral behavior in the infrared. Comparing between the two systems, we find that while the Neptunian satellites have notably higher albedos than the Uranian satellites. The colors (ratio in albedo between any two filters) of the two sets of satellites are similar, but the Uranian moons generally have a higher 1.4-2.1 \micron{} slope, a deeper 3.0 \micron{} feature, and a slightly smaller 3.0-4.6 \micron{} slope than the Neptunian satellites. Notably, Uranus' Puck has nearly the same reflectance at 1.4 \micron{} as at 2.1 \micron{}.

\begin{figure*}
    \centering
    \includegraphics[width=8.5cm]{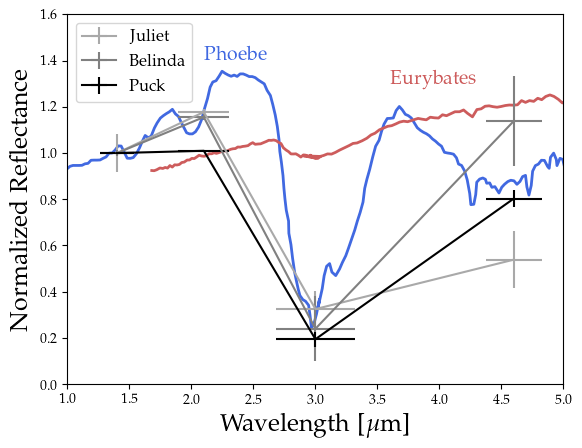}
    \includegraphics[width=8.5cm]{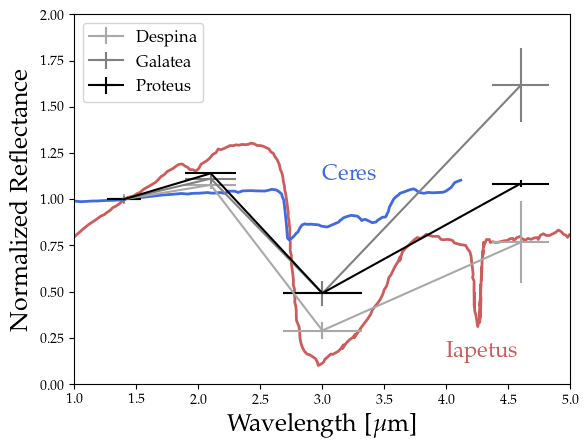}
    \caption{\textit{Left panel:} Comparison of the normalized reflectance of Phoebe from Cassini VIMS data \citep{2019Icar..321..791C}, and Jupiter Trojan Eurybates \citep{2023arXiv231111531W}, to Uranian inner satellites Juliet, Belinda, and Puck. Eurybates, while having a similarly low albedo to the Uranian inner satellites, has a significantly different infrared spectrum, with a much weaker 3.0 \micron{} feature. Phoebe's albedo is similar to the very dark Uranian satellites, and seems to have a similarly deep 3.0 \micron{} feature, with a similar upturn between 1.4 and 2.1 \micron{}. \textit{Right panel:} Comparison of the normalized reflectance of Ceres (from \citealt{life11010009}) and Iapetus from VIMS to Neptunian inner satellites Despina, Galatea, and Proteus. Ceres' 3.0 \micron{} ammonia feature is shallow, and does not match any ice giant inner satellite. The dark part of Iapetus has a spectrum that is compatible with that of Despina, but not Proteus or Galatea.}
    \label{fig:newplot}
\end{figure*}

We note that our measured near-infrared albedos are lower than the HST visible albedos from \citep{2001Icar..151...51K, 2003Icar..162..400K}. Yet other values from previous observations of the satellites, such as from \cite{2003AJ....126.1080D}, obtain near-IR photometry for Proteus with a distinctly lower albedo of near 7\% in the 1.0-2.0 \micron{} range, and find Puck to be higher albedo than Proteus, inverting our results. The HST albedos from \cite{2003AJ....126.1080D} were obtained for the leading hemisphere of Proteus, and the opposite (southern) hemisphere for Puck. Similarly, the old Voyager and \cite{2001Icar..151...51K, 2003Icar..162..400K} HST observations were taken of different orientations of the satellites. Recent ground-based observations by \cite{2023Icar..39115331P} find H-band reflectivities of the Uranian satellites which are comparable to our results when looking at similar phase angles. \cite{2023Icar..39115331P} also find that Belinda and Portia have notably lower reflectivities than Puck in the near-IR, which is a good match to our results. Given the many decades between observations, varying phase angles, and very distinct orientations of these irregularly-shaped objects observed, we do not offer further comparison between our JWST spectrophotometry and past data obtained for the satellites. Future time-series observations of these satellites are necessary to constrain their photometric properties as a function of phase angle and orientation and may help to resolve discrepancies in photometric measurements accumulated over four decades of sparse observations.  

The most prominent feature on the observed satellites is a 3.0 \micron{} absorption that could be attributed to multiple possible materials. Possible matches for this feature include: water ice or hydrated minerals from the O-H stretch at 3.0 \micron{} \citep{1994CCM....42..702B, 1997AREPS..25..243B, 1998ASSL..227..199S}, clays containing ammonia salts such as NH$_4^+$ that has a 3.07 \micron{} feature \citep{2002P&SS...50...11B}, or complex aliphatic and aromatic organics that have a set of C-H stretches just past 3.0 \micron{} \citep{2012JGRE..117.3008H, 2012ApJ...754...75R}. Combinations of these compounds are possible, similar to Kuiper belt objects that have both water ice and hydrocarbons. Most of the satellites show a slight upwards slope between 1.4 and 2.1 \micron{}, a near-IR upturn that is characteristic of less red, dynamically excited KBOs \citep{2012AJ....143..146B, 2012A&A...546A.115H, 2023PSJ.....4...80F}.

Without complete 1-5 \micron{} spectra for the inner satellites, we cannot directly infer their surface composition and determine which of the above hypotheses for surface composition holds. However, we can draw comparisons to known spectra of the gas giant satellites and JWST NIRSpec data of other small outer solar system bodies, in particular KBOs and Neptune Trojans.  In \autoref{fig:compare}, we overlay our JWST NIRCam photometry for the two largest of the inner satellites, Puck and Proteus against the spectrum of Kuiper belt object 2004 EW95 (JWST Program 2418) and Neptune Trojan 2007 VL305 \citep{2023arXiv231003998M}. Both 2004 EW95 and 2007 VL305 have a 3.0 \micron{} feature with a fresnel peak due to water ice, and a CO$_2$ ice feature at 4.25 \micron{}. 2004 EW95 has a notably low albedo and, as a Plutino with a perihelion of 26 AU, experiences similar levels of peak insolation to Proteus. Similarly, the Neptune Trojan 2007 VL305 has likely been at Neptune's L4 point for 4.5 Gyr, thus providing an excellent comparison to Proteus in terms of amount of solar radiation received (the radius and thus albedo for 2007 VL305 is unknown, so we scale the albedo to match that of Proteus). We find that Proteus matches both of these spectra, with a deep 3.0 \micron{} absorption, and similar spectral slopes between 1.4/2.1 \micron{} and 3.0/4.6 \micron{}. However, there are numerous KBOs and Neptune Trojans that do not share these spectral features. Many KBOs with $q<30$ AU have significantly higher albedos than the ones measured for Proteus \citep{2013A&A...557A..60L}, as well as clearly redder spectral slopes in the visible and near-infrared. Additionally, Neptune Trojans with ``very red'' visible spectral slopes, such as 2013 VX30, do not exhibit the 3.0-4.6 \micron{} upturn that Proteus and Puck do (see figure 1 of \citealt{2023arXiv231003998M}).

The Saturnian satellites also offer an important comparison point for our spectrophotometry. Many of the Saturnian satellites have high albedos (Enceladus, Tethys, Iapetus trailing) that are very distinct from the set of Uranian and Neptunian satellites observed, but the darker Phoebe or the leading side of Iapetus are a better point of comparison (see figure 2 of \cite{2012Icar..220.1064F}, also data from \citealt{2012Icar..218..831C,2024arXiv240215698H}). We show data from Cassini VIMS on the Saturnian system from \cite{2019Icar..321..791C} in  \autoref{fig:newplot}. Phoebe and Iapetus both have spectra similar to some of the Neptunian and Uranian moons, largely matching the depth of the 3.0 \micron{} feature and the 1.4-2.1 \micron{} and 3.0-4.6 \micron{} spectral slopes.

Our search for spectral analogues of the small satellites of the ice giants has returned two possible classes of bodies: dark, icy Kuiper belt objects, and two of the Saturnian satellites, Phoebe and Iapetus. However, the 3.0 \micron{} absorption on all the inner satellites is of uncertain provenance, and thus the formation history of moons like Proteus and Puck remains unclear. Although all of the solar system analogues that fit the data for the small satellites have clear water ice absorption features, the low resolution nature of our data precludes us from ruling out ammonia, organics, and hydrated minerals as the drivers of the 3.0 \micron{} feature in our spectrophotometry.

If the inner satellites are similar to KBOs with a 3.0 \micron{} feature due to water ice, it could indicate either that water ice was stable in the ice giant subnebulae at the formation radius of Puck and Proteus \citep{2020ApJ...894..143B}, or that the inner satellites are a mix of collisional material from comets, KBOs, and the primordial inner satellites \citep{1992JGR....9710227C}. While water ice could be present on these objects, it cannot be the dominant component composing all of the small Uranian and Neptunian moons, specifically not the innermost ones. Recent work by \cite{2024Icar..41115957F} has found that the density of Uranian inner satellites increases closer to the planet, consistent with the Roche critical density (see figure 38 of \citealt{2024Icar..41115957F}). The innermost of the inner satellites must be composed of some material with density greater than that of ice, potentially consistent with silicates or organics \citep{1986Sci...233...43S,1989Sci...246.1422S,2001Icar..151...51K,2013ApJ...765L..28T}. Reconciling these varying densities with the consistent 3.0 \micron{} band depth we observe in both systems will require feature observations.

Other possibilities for the 3.0 \micron{} absorption appear less likely. The O-H stretch due to hydrated minerals produces a smaller band depth than seen in our spectrophotometry \citep{1994CCM....42..702B, 2008ClMin..43...35B, 2019Icar..333..243T}. Ammonia is found in the ice giants' atmospheres and on the regular satellites of the Uranian system \citep{2020ApJ...898L..22C}, and its presence on the inner satellites would suggest they consist of subnebular material, as ammonia is not common throughout the Kuiper belt. Ceres has a prominent 3.0 \micron{} feature attributed to ammonia as shown in \autoref{fig:newplot} \citep{life11010009}, but has a much smaller band depth than the small Neptunian and Uranian satellites. The Jupiter Trojan Eurybates is an example of a solar system body with a similar albedo to the inner Uranian moons and a 3.0 \micron{} feature attributed to the O-H stretch mixed with organics \citep{2023arXiv231111531W}. Organics also seem to be a poor match, as the depth of the 3.0 \micron{} absorption on Eurybates is significantly weaker than in our observations.

In order to resolve key questions on formation of satellites around ice giants, new observations that have spectral coverage from 0.6-5.0 \micron{} (with JWST NIRSpec) for the inner satellites of Neptune and Uranus are necessary to pinpoint the surface compositions of these objects. With possible upcoming spacecraft exploration of Uranus in the next several decades, these data prove valuable in target selection and determining the capabilities needed to fully uncover the compositions of the ice giant systems. 

\begin{acknowledgements}
\section*{Acknowledgements}
The data were obtained from the Mikulski Archive for Space Telescopes at the Space Telescope Science Institute, which is operated by the Association of Universities for Research in Astronomy, Inc., under NASA contract NAS 5-03127 for JWST. These observations are associated with program 2739, with the relevant observations linked at the following DOI: \dataset[10.17909/ypkv-m919]{https://doi.org/10.17909/ypkv-m919}. The authors acknowledge PI Klaus Pontoppidan for developing the observing program.
\facility{JWST (NIRCam)}
\software{Astropy \citep{2022ApJ...935..167A}, Photutils \citep{larry_bradley_2023_7946442}, JWST pipeline version 1.11.3 \citep{bushouse_2023_8157276}.}  
\end{acknowledgements}

\bibliography{biblio}{}
\bibliographystyle{aasjournal}
\appendix
We report the fluxes and albedos for the spectrophotometry of the ice giant satellites below. Albedo is calculated as follows:
\begin{equation}
    p = \left(\frac{\text{target flux}}{\text{solar flux}}\right)\div\left(\frac{1 \mathrm{AU}^2}{d_{pj}^2}*\frac{r_{\textrm{moon}}^2}{d_{ps}^2}\right)
\end{equation}
where $d_{ps}$ is the planet-sun distance and $d_{pj}$ is the planet-JWST distance. Radii for objects are from \cite{1986Sci...233...43S, 1989Sci...246.1422S,2001Icar..151...51K, karkoschka2003sizes}, with mean radii used due to the non-sphericity of the satellites.  

\begin{table*}[h!]
\caption{JWST NIRCam Photometry of Neptunian and Uranian Inner Satellites}
\label{tab:table1}
\centering
\begin{tabular}{lllll}
\toprule
 & F140M & F210M & F300M & F460M \\
\midrule
Despina     \\
\hspace{1em} Flux (\textmu Jy) & 7.5 $\pm$ 0.2 & 5.19 $\pm$ 0.10 & 0.76 $\pm$ 0.12  & 0.9 $\pm$ 0.3\\
\hspace{1em} Albedo & 0.092 $\pm$ 0.003 & 0.100 $\pm$ 0.002 & 0.027 $\pm$ 0.004 & 0.07 $\pm$ 0.02 \\
Galatea   \\
\hspace{1em} Flux (\textmu Jy) & 7.68 $\pm$ 0.16 &  5.49 $\pm$ 0.13 & 1.32 $\pm$ 0.19 & 1.9 $\pm$ 0.2\\
\hspace{1em} Albedo & 0.075 $\pm$ 0.002 &  0.084 $\pm$ 0.002 &  0.037 $\pm$ 0.005 & 0.122 $\pm$ 0.015\\
Larissa \\
\hspace{1em} Flux (\textmu Jy) & 9.84 $\pm$ 0.09 & 6.55 $\pm$ 0.10 & 1.42 $\pm$ 0.17 & 1.68 $\pm$ 0.19\\
\hspace{1em} Albedo & 0.0784 $\pm$ 0.0007 & 0.0812 $\pm$ 0.0012 & 0.032 $\pm$ 0.004 & 0.088 $\pm$ 0.010 \\
Proteus     \\
\hspace{1em} Flux (\textmu Jy) & 53.68 $\pm$ 0.10 &  39.34 $\pm$ 0.10 & 9.24 $\pm$ 0.07 & 8.82 $\pm$ 0.13\\
\hspace{1em} Albedo & 0.0913 $\pm$ 0.0002 & 0.1041 $\pm$ 0.0003 & 0.0449 $\pm$ 0.0003 & 0.0990 $\pm$ 0.0015\\
\midrule
Cressida     \\
\hspace{1em} Flux (\textmu Jy) &  6.86 $\pm$ 0.17 & 5.57 $\pm$ 0.13 & 0.49 $\pm$ 0.09 & 1.23 $\pm$ 0.15\\
\hspace{1em} Albedo & 0.0607 $\pm$ 0.0015 & 0.0768 $\pm$ 0.0019 & 0.012 $\pm$ 0.002 & 0.072 $\pm$ 0.009\\
Juliet     \\
\hspace{1em} Flux (\textmu Jy) & 11.2 $\pm$ 0.9 & 8.45 $\pm$ 0.10 & 1.27 $\pm$ 0.31 & 0.9 $\pm$ 0.2\\
\hspace{1em} Albedo & 0.072 $\pm$ 0.006
 & 0.0843 $\pm$ 0.0010 & 0.023 $\pm$ 0.006 & 0.039 $\pm$ 0.009 \\
Portia     \\
\hspace{1em} Flux (\textmu Jy) & 20.20 $\pm$ 0.12 & 13.93 $\pm$ 0.08 &  1.63 $\pm$ 0.11 & 2.48 $\pm$ 0.18\\
\hspace{1em} Albedo & 0.0619 $\pm$ 0.0004 &  0.0664 $\pm$ 0.0004 &  0.0143 $\pm$ 0.0009 & 0.050 $\pm$ 0.004\\
Belinda     \\
\hspace{1em} Flux (\textmu Jy) & 7.51 $\pm$ 0.10 & 5.58 $\pm$ 0.07 & 0.63 $\pm$ 0.37 & 1.3 $\pm$ 0.2\\
\hspace{1em} Albedo &  0.0526 $\pm$ 0.0007 &  0.0607 $\pm$ 0.0007 & 0.013 $\pm$ 0.007 & 0.060 $\pm$ 0.010\\
Puck     \\
\hspace{1em} Flux (\textmu Jy) & 33.03 $\pm$ 0.11 & 21.44 $\pm$ 0.10 & 2.2 $\pm$ 0.4 & 4.02 $\pm$ 0.19\\
\hspace{1em} Albedo & 0.0713 $\pm$ 0.0002 & 0.0721 $\pm$ 0.0004 &  0.014 $\pm$ 0.002 & 0.057 $\pm$ 0.003\\
\bottomrule
\end{tabular}
\end{table*}

\end{document}